\documentclass[preprint,aps,showpacs,nofootinbib,tightenlines]{revtex4}
\usepackage{amsmath}
\usepackage{amssymb}
\usepackage{epsfig}
\usepackage{graphicx}
\begin{document}
\preprint{XZNU-PHY-TH-11-01}

\newcommand{\beq}{\begin{eqnarray}}
\newcommand{\eeq}{\end{eqnarray}}
\newcommand{\non}{\nonumber\\ }

\newcommand{\acp}{{\cal A}_{CP}}
\newcommand{\etap}{\eta^{(\prime)} }
\newcommand{\etapr}{\eta^\prime }
\newcommand{\jpsi}{ J/\Psi }
\newcommand{\kst} {K_0^{*}(1430)}
\newcommand{\kstb}{\overline{K}_0^{*}(1430)}
\newcommand{\ks} {K_0^*}
\newcommand{\ksb} {\overline{K}_0^*}

\newcommand{\psl}{ P \hspace{-2.8truemm}/ }
\newcommand{\nsl}{ n \hspace{-2.2truemm}/ }
\newcommand{\vsl}{ v \hspace{-2.2truemm}/ }
\newcommand{\epsl}{\epsilon \hspace{-1.8truemm}/\,  }

\def \epjc{ Eur. Phys. J. C }
\def \jpg{  J. Phys. G }
\def \npb{  Nucl. Phys. B }
\def \npps { Nucl. Phys. B (Proc. Suppl.) }
\def \plb{  Phys. Lett. B }
\def \pr{  Phys. Rep. }
\def \prd{  Phys. Rev. D }
\def \prl{  Phys. Rev. Lett.  }
\def \zpc{  Z. Phys. C  }
\def \jhep{ J. High Energy Phys.  }
\def \ijmpa { Int. J. Mod. Phys. A }
\def \cpc{ Chin. Phys. C }
\def \ctp{ Commun. Theor. Phys. }
\def \rmp{ Rev. Mod. Phys. }
\def \ppnp{ Prog. Part. $\&$ Nucl. Phys. }

\title{ Charmless hadronic $B_{q} \to \kst \kstb$ decays in the pQCD approach }
\author{
Xin~Liu$^{1,2}$\footnote{liuxin.physics@gmail.com},
Zhen-Jun~Xiao$^2$\footnote{xiaozhenjun@njnu.edu.cn},
 and
Zhi-Tian~Zou$^3$\footnote{zouzt@ihep.ac.cn}}
\affiliation{
$^1$ Department of Physics and Institute of Theoretical Physics,
Xuzhou Normal University, Xuzhou,\;\; \\ Jiangsu 221116, People's Republic of China\\
$^2$ Department of Physics and Institute of Theoretical Physics,
Nanjing Normal University, Nanjing, Jiangsu 210046, People's Republic of China\\
$^3$ Institute of High Energy Physics, CAS, P.O.Box 918(4)
Beijing 100049, People's Republic of China }
\date{\today}
\begin{abstract}
\bigskip
Based on the assumption of two-quark structure of the scalar $K_0^*(1430)$, the CP-averaged
branching ratios(BRs) and CP-violating asymmetries of charmless hadronic
$B_q(q=u,d,s,c) \to \kst \kstb$ decays are studied in the standard model(SM) by employing the
perturbative QCD(pQCD) factorization approach.
Our predictions are the following:
(1) the CP-averaged BRs for $B_{q} \to \kst \kstb$ decays in both scenarios
vary in the range of $10^{-6} \sim 10^{-4}$ in the SM;
(2) the magnitudes of $\acp^{\rm dir}(B_u \to {\kst}^+ {\kstb}^0)$ and
$\acp^{\rm dir}(B_{d,s} \to {\kst}^+ {\kst}^-)$ in Scenario 1 are much larger than
those in Scenario 2 correspondingly;
(3) there are no direct CP violations in $B_{d,s} \to  {\kst}^0 {\kstb}^0$ and
$B_c \to {\kst}^+ {\kstb}^0$ decays
in the SM because of the pure penguin and tree topology, respectively.
A measurement of our pQCD predictions at the predicted level will
favor the $q\bar q$ structure of the scalar $\kst$ and help us to understand its physical
properties and the involved QCD dynamics.
\end{abstract}


\maketitle

\section{Introduction}

The exclusive non-leptonic weak decays of $B_q$ mesons ($q=(u,d,s,c)$) provide not
only good opportunities for testing the Standard Model (SM) but
also excellent places for probing different new physics scenarios
beyond the SM. Very recently, the Belle collaboration has reported a
preliminary upper limit on branching ratio of charmless hadronic
$B^0 \to {\kst}^0 {\kstb}^0$\footnote{In the following section,
we will adopt $\ks$ to denote ${\kst}^+$ and
${\kst}^0$, and $\ksb$ to stand for ${\kst}^-$ and ${\kstb}^0$ for convenience,
respectively, unless otherwise stated. }
decay~\cite{Chiang10:k0k0b,Chen11:belle,Asner10:hfag}:
\beq
Br(B^0 \to {\ks}^0 {\ksb}^0) &=&
3.21^{+2.89+2.31}_{-2.85-2.32} \times 10^{-6}\;\; {\rm or}\;\; <8.4 \times 10^{-6}\; \;  {\rm at} \; \;
90\%\; {\rm C.L.}
\eeq
where the result is fitted for decay mode with final states
$K^+\pi^-K^-\pi^+$ and $Br({\ks}^0 \to K^+ \pi^-) \approx 66.7\%$.
This measurement will be improved soon with the ongoing Large
Hadron Collider(LHC) experiments at CERN. At LHC experiments, the
$b$ hadrons such as $B_u$, $B_d$, $B_s$, $B_c$, even $\Lambda_b$
can be accessed easily. Particularly, the $B_c$ meson could be
produced abundantly, which will make a new realm to test the SM, study the
heavy flavor dynamics, and explore the involved perturbative and nonperturbative
QCD dynamics~\cite{Brambilla04:bcreview}.

In the naive quark model, $\ks$ is a $p$-wave scalar($1^3P_0$) particle with quantum
number $J^{PC}=0^{++}$. Lattice
calculations~\cite{Mathur06:lattice} on the masses of $\ks$ and
$a_0(1450)$ indicate a good SU(3) symmetry for the scalar
sector, while the latter has been confirmed to be a $\bar q q$ meson in
lattice calculations~\cite{Mathur06:lattice,Kim97-Burch06:lattice,Bardeen02:lattice,
Kunihiro04:lattice,Prelovsek04:lattice}. Recently, Cheng, Chua,
and Yang proposed two possible scenarios based on the assumption of two-quark structure to describe this light
scalar $\ks$ in the QCD sum rule method~\cite{Cheng06:B2SP}: the first excited
state in scenario 1(S1) or the lowest lying state in scenario
2(S2), and made extensive studies and interesting analyses
phenomenologically on charmless hadronic $B \to \ks (P, V)$
(Here, $P$ and $V$ stand for the light pseudoscalar and vector
mesons, respectively) decays to implicate its
physical properties in the QCD factorization(QCDF) approach~\cite{Beneke99:qcdf,Du02:qcdf}.
Moreover, the people also made relevant investigations on $\ks$ with other approaches/methods
in hadronic $B$ meson decays~
\cite{Chen07:b2kphi,Shen07:b2kpi,Kim10:b2kphi,Liu10:b2k0s,Liu10:bc2s}.

It is well known that the key point of the theoretical calculation for the
charmless hadronic $B_q$ meson decays is how to calculate the hadronic
matrix element(HME) reliably.
So far, many theoretical
approaches/methods, such as naive factorization
assumption~\cite{Wirbel85:nfa}, generalized factorization
approach~\cite{Ali98:gnf}, QCDF, soft-collinear
effective theory(SCET)~\cite{Bauer04:scet}, and perturbative
QCD(pQCD) approach~\cite{Li01:kpi,Lu01:pipi,Li03:ppnp}, are
developed to make effective evaluations of HME and interpret the
existing rich data.
Up to now, the pQCD approach has become one of the most popular methods
due to its unique features~\cite{Li09:review}.
The annihilation diagrams, for example, can be evaluated here. While
the strong phase for generating CP violation~\cite{Chay08:complexanni} in the pQCD factorization
approach, is rather different from that as claimed in SCET~\cite{Arnesen08:anni-scet}.

Motivated by the above observations on both experiment and theory aspects, we will
investigate 
the charmless hadronic $B_q \to \ks \ksb$\footnote{Hereafter, for the sake of simplicity,
we will adopt $B$ to denote the $B_u$ and $B_d$ mesons, unless otherwise
stated. } decays with $q=u,\;d,\;s,\;{\rm and}\;c$ by employing
the low energy effective Hamiltonian~\cite{Buras96:weak} and the
perturbative QCD(pQCD) factorization approach in this work.
We here not only calculate the usual factorizable contributions, but also evaluate
the nonfactorizable and the annihilation type contributions
theoretically. We will predict the physical observables such as
CP-averaged branching ratios(BRs) and CP-violating asymmetries in the
considered decays.
The large BRs and CP violations in the relevant considered decay channels will
play an important role in exploring the physical properties of scalar $\ks$.
Furthermore, the pure annihilation processes $B_d \to {\ks}^+ {\ks}^-$ and $B_c \to
{\ks}^+ {\ksb}^0$ could provide interesting information
to explore the underlying decay mechanism of the weak annihilation decays.

The paper is organized as follows. In Sec.~\ref{sec:1}, we present
the theoretical framework on the low energy effective Hamiltonian
and formalism of the pQCD approach.
Then we perform the perturbative calculations for the considered
$B_q \to \ks \ksb$ decay channels with pQCD approach in Sec.~\ref{sec:2}.
The analytic formulas of the decay amplitudes for all the considered
 modes are also collected in this section.
The numerical results and phenomenological analysis are given in
Sec.~\ref{sec:3}. Finally, Sec.~\ref{sec:sum} contains the main
conclusions and a short summary.

\section{Theoretical framework }\label{sec:1}

For the considered decays,
the related weak effective
Hamiltonian $H_{{\rm eff}}$~\cite{Buras96:weak} can be written as
\begin{equation}
H_{\rm eff}\, =\, {G_F\over\sqrt{2}}
\sum_{Q=u,c} V^*_{Qb}V_{QD}\left[C_1(\mu)O_1^{(Q)}(\mu)
+C_2(\mu)O_2^{(Q)}(\mu)+ \sum_{i=3}^{10}C_i(\mu)O_i(\mu)\right]+ {\rm H.c.}\;,
\label{eq:heff}
\end{equation}
with the Fermi constant $G_F=1.16639\times 10^{-5}{\rm
GeV}^{-2}$, Cabibbo-Kobayashi-Maskawa(CKM) matrix elements $V$,
light down type quarks $D = d, s$,
and Wilson coefficients $C_i(\mu)$ at the renormalization scale
$\mu$. The local four-quark
operators $O_i(i=1,\cdots,10)$ are written as
\begin{enumerate}
\item[]{(1) Current-current(tree) operators}
\begin{eqnarray}
{\renewcommand\arraystretch{1.5}
\begin{array}{ll}
\displaystyle
O_1^{(Q)}\, =\,
(\bar{D}_\alpha Q_\beta)_{V-A}(\bar{Q}_\beta b_\alpha)_{V-A}\;,
& \displaystyle
O_2^{(Q)}\, =\, (\bar{D}_\alpha Q_\alpha)_{V-A}(\bar{Q}_\beta b_\beta)_{V-A}\;;
\end{array}}
\label{eq:operators-1}
\end{eqnarray}

\item[]{(2) QCD penguin operators}
\begin{eqnarray}
{\renewcommand\arraystretch{1.5}
\begin{array}{ll}
\displaystyle
O_3\, =\, (\bar{D}_\alpha b_\alpha)_{V-A}\sum_{q'}(\bar{q}'_\beta q'_\beta)_{V-A}\;,
& \displaystyle
O_4\, =\, (\bar{D}_\alpha b_\beta)_{V-A}\sum_{q'}(\bar{q}'_\beta q'_\alpha)_{V-A}\;,
\\
\displaystyle
O_5\, =\, (\bar{D}_\alpha b_\alpha)_{V-A}\sum_{q'}(\bar{q}'_\beta q'_\beta)_{V+A}\;,
& \displaystyle
O_6\, =\, (\bar{D}_\alpha b_\beta)_{V-A}\sum_{q'}(\bar{q}'_\beta q'_\alpha)_{V+A}\;;
\end{array}}
\label{eq:operators-2}
\end{eqnarray}

\item[]{(3) Electroweak penguin operators}
\begin{eqnarray}
{\renewcommand\arraystretch{1.5}
\begin{array}{ll}
\displaystyle
O_7\, =\,
\frac{3}{2}(\bar{D}_\alpha b_\alpha)_{V-A}\sum_{q'}e_{q'}(\bar{q}'_\beta q'_\beta)_{V+A}\;,
& \displaystyle
O_8\, =\,
\frac{3}{2}(\bar{D}_\alpha b_\beta)_{V-A}\sum_{q'}e_{q'}(\bar{q}'_\beta q'_\alpha)_{V+A}\;,
\\
\displaystyle
O_9\, =\,
\frac{3}{2}(\bar{D}_\alpha b_\alpha)_{V-A}\sum_{q'}e_{q'}(\bar{q}'_\beta q'_\beta)_{V-A}\;,
& \displaystyle
O_{10}\, =\,
\frac{3}{2}(\bar{D}_\alpha b_\beta)_{V-A}\sum_{q'}e_{q'}(\bar{q}'_\beta q'_\alpha)_{V-A}\;.
\end{array}}
\label{eq:operators-3}
\end{eqnarray}
\end{enumerate}
with the color indices $\alpha, \ \beta$ and the notations
$(\bar{q}'q')_{V\pm A} = \bar q' \gamma_\mu (1\pm \gamma_5)q'$.
The index $q'$ in the summation of the above operators runs
through $u,\;d,\;s$, $c$, and $b$.
The standard combinations $a_i$ of Wilson coefficients are defined as follows,
  \beq
a_1&=& C_2 + \frac{C_1}{3}\;, \qquad  a_2 = C_1 + \frac{C_2}{3}\;,
\qquad  a_i = C_i + \frac{C_{i \pm 1}}{3}(i=3 - 10) \;.
  \eeq
where the upper(lower) sign applies, when $i$ is odd(even).
Since we work in the leading order[${\cal O}(\alpha_s)$] of the pQCD
approach, it is consistent to use the leading order Wilson coefficients,
although the
next-to-leading order calculations already exist in the
literature~\cite{Buras96:weak}. This is the consistent way to
cancel the explicit $\mu$ dependence in the theoretical formulae.
For the renormalization group evolution of the Wilson coefficients
from higher scale to lower scale, we use the formulas as given in
Ref.~\cite{Lu01:pipi} directly.

The basic idea of the pQCD approach is that it takes into
account the transverse momentum ${\bf k}_T$ of the valence quarks
in the calculation of the hadronic matrix elements.
The $B_q$ meson transition form factors, and the spectator and
annihilation contributions are then all calculable in the framework
of the ${\bf k}_T$ factorization. In the pQCD approach, a $B_q \to M_2 M_3$
decay amplitude is factorized into the convolution of the six-quark
hard kernel($H$), the jet function($J$) and the Sudakov factor($S$) with the
bound-state wave functions($\Phi$) as follows,
 \begin{eqnarray}
{\cal A}(B_q \to M_2 M_3)=\Phi_{B_q} \otimes H \otimes J \otimes S
\otimes \Phi_{M_2} \otimes \Phi_{M_3}\;, \label{eq:sixquarks}
\end{eqnarray}
The jet function $J$ comes from the threshold resummation, which
exhibits suppression in the small $x$(quark momentum fraction) region\cite{Li02:threshold}. The
Sudakov factor $S$ comes from the ${\bf k}_T$ resummation, which
exhibits suppression in the small ${\bf k}_T$ region\cite{Botts89:ktfact,Li92:sudakov}.
Therefore, these resummation
effects guarantee the removal of the endpoint singularities.

In the practical applications to heavy $B_q$ meson decays, the
decay amplitude of Eq.~(\ref{eq:sixquarks}) in the pQCD approach
can be conceptually written as\footnote{$J$($S$), organizing
double logarithms in the hard kernel (meson wave functions), is
hidden in $H$ (the three meson states).}, \beq {\cal A}(B_q \to
M_2 M_3) &\sim &\int\!\! d^4k_1d^4k_2d^4k_3 \mathrm{Tr} \left [
C(t) \Phi_{B_q}(k_1) \Phi_{M_2}(k_2) \Phi_{M_3}(k_3)
H(k_1,k_2,k_3, t) \right ],\label{eq:a1} \eeq where $k_i$'s are
momenta of light quarks included in each mesons, and $\mathrm{Tr}$
denotes the trace over Dirac and color indices. $C(t)$ is the
Wilson coefficient which results from the radiative corrections at
short distance. In the above convolution, $C(t)$ includes the
harder dynamics at larger scale than $m_{B_q}$ scale and describes
the evolution of local $4$-Fermi operators from $m_W$ (the $W$
boson mass) down to $t\sim\mathcal{O}(\sqrt{\Lambda_{\rm QCD}
m_{B_q}})$ scale, where $\Lambda_{\rm QCD}$ is the hadronic scale.
The function $H(k_1,k_2,k_3,t)$ describes the four quark operator
and the spectator quark connected by
 a hard gluon whose $q^2$ is in the order
of $\Lambda_{\rm QCD} m_{B_q}$, and includes the
$\mathcal{O}(\sqrt{\Lambda_{\rm QCD} m_{B_q}})$ hard dynamics.
Therefore, this hard part $H$ can be perturbatively calculated.
The function $\Phi_M$ is the wave function which describes
hadronization of the quark and anti-quark to the meson $M$, which
is independent of the specific processes and usually determined by
employing nonperturbative QCD techniques or other well measured
processes.

Since the b quark is rather heavy, we work in the frame with the
$B_q$ meson at rest for simplicity.
Throughout this paper, we will use light-cone
coordinate $(P^+, P^-, {\bf P}_T)$ to describe the meson's momenta with the definitions
$P^{\pm}=\frac{p_0 \pm p_3}{\sqrt{2}}$ and ${\bf P}_T=(p_1,p_2)$.
For the charmless hadronic $B_u \to {\ks}^+ {\ksb}^0$
decay, for example, assuming that the ${\ks}^+$ (${\ksb}^0$) meson moves in the plus
(minus) $z$ direction carrying the momentum $P_2$ ($P_3$). Then the two final state meson momenta can be
written as
\beq
     P_1&=&\frac{m_{B}}{\sqrt{2}}(1,1,{\bf 0}_T)\;, \quad
     P_2 =\frac{m_{B}}{\sqrt{2}} (1-r_3^2,r_2^2,{\bf 0}_T)\;, \quad
     P_3 =\frac{m_{B}}{\sqrt{2}} (r_3^2,1-r_2^2,{\bf 0}_T)\;,
\eeq
respectively, where $r_2=m_{\ks}/m_{B}$ and
$r_3=m_{\ksb}/m_{B}$.
Putting the (light-)
 quark momenta in $B$, $\ks$ and $\ksb$ mesons as $k_1$,
$k_2$, and $k_3$, respectively, we can choose
\beq
k_1 = (x_1P_1^+,0,{\bf k}_{1T}), \quad k_2 = (x_2 P_2^+,0,{\bf k}_{2T}), \quad
k_3 = (0, x_3 P_3^-,{\bf k}_{3T})\;;
\eeq
Then, for $B_u \to {\ks}^+ {\ksb}^0$
decay, the
integration over $k_1^-$, $k_2^-$, and $k_3^+$
will conceptually lead to the decay amplitude in the pQCD approach,
\beq
{\cal A}(B_u \to {\ks}^+ {\ksb}^0) &\sim &\int\!\! d x_1 d x_2 d x_3 b_1
d b_1 b_2 d b_2 b_3 d b_3 \non && \times \mathrm{Tr} \left [ C(t)
\Phi_{B_u}(x_1,b_1) \Phi_{\ks}(x_2,b_2) \right. \non && \left.
\times \Phi_{\ksb}(x_3, b_3) H(x_i,
b_i, t) S_t(x_i)\, e^{-S(t)} \right ]\;.
\label{eq:a2}
\eeq
where $b_i$ is the conjugate space coordinate of ${\bf k}_{iT}$, and $t$ is the
largest energy scale in function $H(x_i,b_i,t)$. The large
logarithms $\ln (m_W/t)$ are included in the Wilson coefficients
$C(t)$. The large double logarithms ($\ln^2 x_i$) are summed by the
threshold resummation~\cite{Li02:threshold}, and they lead to the jet function
$S_t(x_i)$ which smears the end-point singularities on $x_i$. The
last term, $e^{-S(t)}$, is the Sudakov factor which suppresses
the soft dynamics effectively~\cite{Li98:soft}. Thus it makes the
perturbative calculation of the hard part $H$ applicable at
intermediate scale, i.e., $m_{B}$ scale. We will calculate
analytically the function $H(x_i,b_i,t)$ for the considered decays
at leading order in $\alpha_s$ expansion and give the convoluted
amplitudes in next section.

In the resummation procedures, the heavy $B_q$ meson
is treated as a heavy-light system( In the present work, the $B_c$ meson can also be viewed
as a heavy-light system although $c$ is the known heavy flavor quark.
).
In principle there are two Lorentz structures in the $B_q$ meson's wave function.
One should consider
both of them in calculations. However, since the contribution
induced by one Lorentz structure is numerically small~\cite{Li03:ppnp,Lu03:form,Ali07:bsnd},
and can be neglected approximately, we only consider
the contribution from the first Lorentz structure
 \beq
 \Phi_{B_q}(k) &=& \frac{i}{\sqrt{2 N_c}} \biggl[ (\psl + m_{B_q}) \gamma_5 \phi_{B_q} (k) \biggr]_{\alpha\beta} \;,
 \eeq
where $P(m)$ is the momentum(mass) of the $B_q$ meson, $k$ is the momentum carried by the
light quark in $B_q$ meson, and $\phi_{B_q}$ is the corresponding distribution amplitude,
respectively. In the next section, we will see that the hard part is always
independent of one of the $k^+$ and/or $k^-$, if we make the
approximations shown in the next section. The $B_q$ meson distribution amplitude
$\phi_{B_q}(k)$ is then the function of variables $k^-$(or $k^+$) and ${\bf k}_T$ only,
 \beq
 \phi_{B_q}(k^-, {\bf k}_T) &=& \int \frac{d^4k}{2 \pi} \phi_{B_q} (k^+, k^-,{\bf k}_T)\;;
 \eeq
The transverse momentum ${\bf k}_T$ is usually conveniently converted to the $b$ space parameter
by Fourier transformation.

The light-cone wave function of the light scalar $K^*_0$ has been investigated in the QCD sum rule
method as~\cite{Cheng06:B2SP}
 \beq
 \Phi_{K_0^*}(x) &=& \frac{i}{\sqrt{2 N_c}} \biggl\{\psl\, \phi_{K_0^*}(x) +
 m_{K_0^*}\, \phi_{K_0^*}^S(x) + m_{K_0^*} (\nsl \vsl - 1)\, \phi_{K_0^*}^T(x) \biggr\}_{\alpha\beta}\;.
 \eeq
where $\phi_{K_0^*}$ and $\phi_{K_0^*}^{S,T}$, and $m_{K_0^*}$
are the leading twist and twist-3 distribution amplitudes, and mass
of the scalar $K_0^*$ meson, respectively, while $x$ denotes the momentum
fraction carried by quark in the meson, and $n=(1,0, {\bf 0}_T)$
and $v=(0,1,{\bf 0}_T)$ are the dimensionless light-like unit vectors.

\section{perturbative calculations in the pQCD approach} \label{sec:2}

In the following, we will present the analytic factorization formulas for
charmless hadronic $B_q \to \ks \ksb$ decays in the pQCD approach. Apart from the
factorizable and nonfactorizable spectator diagrams, we can also
calculate analytically the annihilation-type ones with no endpoint
singularity by employing the pQCD approach. We will adopt $F$ and
$M$ to stand for the contributions of factorizable and
nonfactorizable diagrams from $(V-A)(V-A)$ operators, $F^{P1}$ and
$M^{P1}$ to stand for the contribution from $(V-A)(V+A)$
operators, and $F^{P2}$ and $M^{P2}$ to stand for the contribution
from $(S-P)(S+P)$ operators which result from the Fierz
transformation of the $(V-A)(V+A)$ operators.

\begin{figure}[t,b]
\vspace{-0.5cm} \centerline{\epsfxsize=16 cm \epsffile{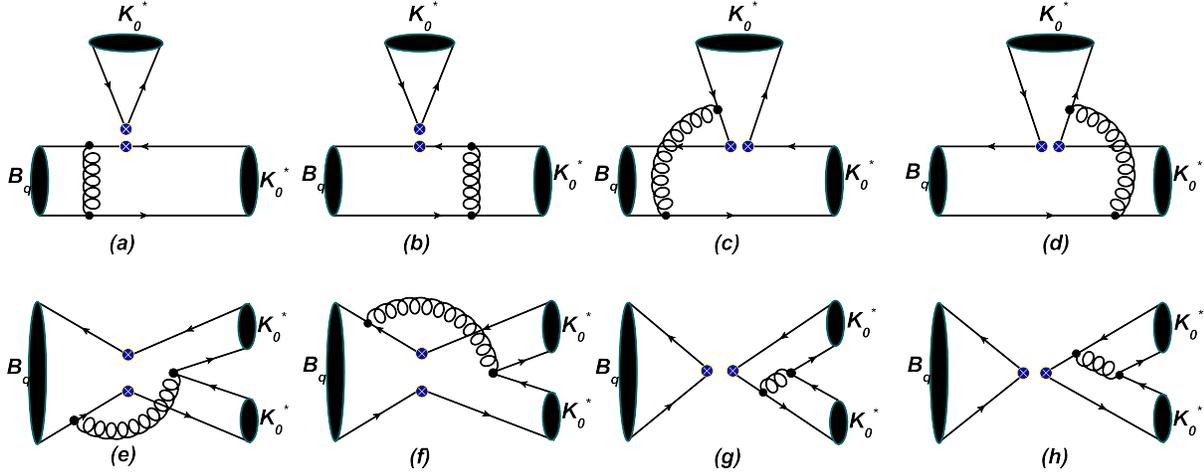}}
\vspace{0.2cm} \caption{Typical Feynman diagrams for charmless
hadronic $B_q \to \ks \ks$ 
decays at leading
order, where $q= u, d, s$, and $c$, respectively. }
 \label{fig:fig1}
\end{figure}

\subsection{$B/B_s \to \ks  \ksb$ decays}

As illustrated in Fig.~\ref{fig:fig1}, when $q$ is $u$, $d$ or $s$,
all eight  types of diagrams may contribute to the $B_{(s)} \to \ks \ksb$ decays.
We firstly calculate the usual factorizable spectator($fs$) diagrams (a) and
(b), in which one can factor out the form factors $B \to \ks$ and $B_s \to \ksb$.
The corresponding Feynman decay amplitudes are given as follows,
\begin{enumerate}
\item[]{(i) $(V-A)(V-A)$ operators:}
 \beq
F_{fs}&=& -8 \pi C_F f_{S}
m_{B_{(s)}}^2\int_0^1 d x_{1} dx_{3}\, \int_{0}^{\infty} b_1 db_1 b_3
db_3\,
\non & & 
\times \phi_{B_{(s)}}(x_1,b_1)\left\{ \left[(1+x_3 )\phi_{S}(x_3) + r_S (1-2
x_3) (\phi^S_{S}(x_3)+\phi^T_{S}(x_3))\right]  \right.\non & &
\left. \times h_{fs}(x_1,x_3,b_1,b_3)E_{fs}(t_a)+ 2\; r_S\; \phi^S_{S} (x_3)\;
h_{fs}(x_3,x_1,b_3,b_1) \;E_{fs}(t_b)
\right\}\;, \label{eq:b-fs}
 \eeq

\item[]{(ii) $(V-A)(V+A)$ operators:}
\beq
F_{fs}^{P1}&=& F_{fs}\;, \label{eq:b-fs-p1}
\eeq

\item[]{(iii) $(S-P)(S+P)$ operators:}
 \beq
 F_{fs}^{P2}&=& 16\pi C_F {\bar f}_{S}
m_{B_{(s)}}^2 r_S  \int_0^1 d x_{1} dx_{3}\, \int_{0}^{\infty} b_1 db_1 b_3
db_3\,  \non &&
\times \phi_{B_{(s)}}(x_1,b_1)\left\{
\left[\phi_{S}(x_3)+ r_S [(2+x_3) \phi^S_{S}(x_3)- x_3
\phi^T_{S}(x_3)]\right] \right.
\non && \left. \times h_{fs}(x_1,x_3,b_1,b_3)E_{fs}(t_a)+ 2\; r_S\;
\phi^S_{S} (x_3) \; h_{fs}(x_3,x_1,b_3,b_1)\;
E_{fs}(t_b)\right\} \label{eq:b-fs-p2}\;;
\eeq
\end{enumerate}
where $r_S=m_{S}/m_B$(Hereafter, for simplicity, we will use $S$
to denote $\ks$ and its charge conjugation $\ksb$ in the explicit
expressions of factorization formulas.) and $C_F=4/3$ is a color
factor. The convolution functions $E_i$, the factorization hard
scales $t_i$, and the hard functions $h_i$ can be referred to
Ref.~\cite{Xiao:pqcd}.

For the non-factorizable spectator($nfs$) diagrams 1(c) and 1(d), the corresponding
decay amplitudes can be written as
\begin{enumerate}
\item[]{(i) $(V-A)(V-A)$ operators:}
\beq
M_{nfs}&=&-
\frac{32}{\sqrt{6}}\pi C_F m_{B_{(s)}}^2 \int_{0}^{1}d x_{1}d x_{2}\,d
x_{3}\,\int_{0}^{\infty} b_1d b_1 b_2d b_2\,
\non & & \times
\phi_{B_{(s)}}(x_1,b_1)
\phi_{S}(x_2)\left\{ \left[(1-x_2)\phi_{S}(x_3) - r_S x_3 (\phi^S_{S}(x_3)-
\phi^T_{S}(x_3))\right]\right.\non & & \left.
\times E_{nfs}(t_c)h_{nfs}^c(x_1,x_2,x_3,b_1,b_2)
-h_{nfs}^d(x_1,x_2,x_3,b_1,b_2)\right.\non & & \left. \times
\left[(x_2+x_3)\phi_{S}(x_3) - r_S x_3 (\phi^S_{S}(x_3)+
\phi^T_{S}(x_3))\right] E_{nfs}(t_d)
\right\} \label{eq:b-nfs}\; ,
\eeq

\item[]{(ii) $(V-A)(V+A)$ operators:}
\beq
M_{nfs}^{P1}&=&\frac{32}{\sqrt{6}}\pi C_F m_{B_{(s)}}^2 \int_{0}^{1}d
x_{1}d x_{2}\,d x_{3}\,\int_{0}^{\infty} b_1d b_1 b_2d b_2\, \non && \times
\phi_{B_{(s)}}(x_1,b_1) r_{S}  \left\{
\left[(1-x_2)\phi_{S}(x_3) (\phi_{S}^S(x_2)
+\phi_{S}^T(x_2)) \right.\right.\non
 & & \left.\left. - r_S\left(\phi_{S}^S(x_2)[(x_2-x_3-1)
\phi^S_{S}(x_3)-(x_2+x_3-1)\phi^T_{S}(x_3)]\right.
\right.\right.\non && \left.\left.\left. +\phi_{S}^T(x_2)
[(x_2+x_3-1)\phi_{S}^S(x_3)+(1-x_2+x_3)\phi_{S}^T]\right)\right]
E_{nfs}(t_c)\right.\non &&\left.
\times h_{nfs}^c(x_1,x_2,x_3,b_1,b_2) -h_{nfs}^d(x_1,x_2,x_3,b_1,b_2)E_{nfs}(t_d)
\right.\non &&\left. \times
\left[x_2(\phi_{S}^S(x_2)-\phi_{S}^T(x_2))\phi_{S}(x_3)+ r_S
(x_2 (\phi_{S}^S(x_2)-\phi_{S}^T(x_2))\right.\right. \non &&\left.\left.
\times (\phi^S_{S}(x_3)-
\phi^T_{S}(x_3)) + x_3
(\phi_{S}^S(x_2)+\phi_{S}^T(x_2))(\phi^S_{S}(x_3)+
\phi^T_{S}(x_3)))\right]
\right\} \label{eq:b-nfs-p1}\;,
\eeq

\item[]{(iii) $(S-P)(S+P)$ operators:}
\beq
 M_{nfs}^{P2}&=& \frac{32}{\sqrt{6}}\pi C_F m_{B_{(s)}}^2
\int_{0}^{1}d x_{1}d x_{2}\,d x_{3}\,\int_{0}^{\infty} b_1d b_1 b_2d
b_2\, \phi_{B_{(s)}}(x_1,b_1) 
\non & &  
\times  \phi_{S}(x_2)\left\{\left[(x_2-x_3-1)\phi_{S}(x_3) +
r_S x_3 (\phi^S_{S}(x_3)+
\phi^T_{S}(x_3))\right]\right. \non
& &\left. \times E_{nfs}(t_c)h_{nfs}^c(x_1,x_2,x_3,b_1,b_2)
+h_{nfs}^d(x_1,x_2,x_3,b_1,b_2)
\right. \non
& &\left.   \times\left[ x_2 \phi_{S}(x_3) - r_S x_3
(\phi^S_{S}(x_3)- \phi^T_{S}(x_3))\right]
E_{nfs}(t_d) \right\} \label{eq:b-nfs-p2}\;;
\eeq
\end{enumerate}
In the above three formulas, i.e., Eqs.~(\ref{eq:b-nfs})-(\ref{eq:b-nfs-p2}),
one can find that there exist cancelations between the contributions of the
two diagrams in Fig.~\ref{fig:fig1}(c) and \ref{fig:fig1}(d).

The Feynman diagrams shown in Fig.~\ref{fig:fig1}(e) and \ref{fig:fig1}(f) are
the non-factorizable annihilation($nfa$) ones, whose contributions are
\begin{enumerate}
\item[]{(i) $(V-A)(V-A)$ operators:}
\beq
M_{nfa}&=& -\frac{32}{\sqrt{6}}\pi C_F m_{B_{(s)}}^2
\int_{0}^{1}d x_{1}d x_{2}\,d x_{3}\,\int_{0}^{\infty} b_1d b_1 b_2d
b_2\,  
\non & &
 \times \phi_{B_{(s)}}(x_1,b_1)\left\{
\left[(1-x_3)\phi_{S}(x_2)\phi_{S}(x_3)- r_S  r_S \left(\phi_{S}^S(x_2)
\right.\right.\right.
\non && \left.\left.\left.
\times [(1+x_2-x_3)
\phi^S_{S}(x_3) -
(1-x_2-x_3)\phi^T_{S}(x_3)]+\phi_{S}^T(x_2)\right.\right.\right.
\non && \left.\left.\left.
\times [(1-x_2-x_3)\phi_{S}^S(x_3)-(1+x_2-x_3)\phi_{S}^T(x_3)]
\right)\right] E_{nfa}(t_e)  \right.\non
& & \left.\times h_{nfa}^e(x_1,x_2,x_3,b_1,b_2)-E_{nfa}(t_f)
h_{nfa}^f(x_1,x_2,x_3,b_1,b_2) \right. \non && \left.
\times\left[x_2 \phi_{S}(x_2)\phi_{S}(x_3)+ r_S r_S
\left(\phi_{S}^S(x_2)[(x_3-x_2-3) \phi^S_{S}(x_3)\right.\right. \right.\non &&
\left. \left.\left.-(1-x_2-x_3)
\phi^T_{S}(x_3)]+\phi_{S}^T(x_2)
[(1-x_2-x_3)\phi_{S}^S(x_3)\right.\right. \right.\non &&
\left. \left.\left.
-(1-x_2+x_3)\phi_{S}^T(x_3)]\right)\right]
 \right\} \label{eq:b-nfa}\;,
\eeq

\item[]{(ii) $(V-A)(V+A)$ operators:}
\beq
M_{nfa}^{P1}&=& \frac{32}{\sqrt{6}}\pi C_F m_{B_{(s)}}^2 \int_{0}^{1}d
x_{1}d x_{2}\,d x_{3}\,\int_{0}^{\infty} b_1d b_1 b_2d b_2\,
 \non
 & & 
\times \phi_{B_{(s)}}(x_1,b_1)\left\{ \left[ r_S
x_2 \phi_{S}(x_3)(\phi_{S}^S(x_2)+\phi_{S}^T(x_2)) + r_S (1 -x_3)
\right.\right.
 \non
 & & \left.\left.
\times \phi_{S}(x_2)(\phi^S_{S}(x_3)- \phi^T_{S}(x_3))\right]
 E_{nfa}(t_e)h_{nfa}^e(x_1,x_2,x_3,b_1,b_2)\right. \non & & \left.
 + \left[ r_S
(2-x_2)(\phi_{S}^S(x_2)+\phi_{S}^T(x_2))
 \phi_{S}(x_3)+r_S (1+x_3)\right.\right.\non && \left.\left.\times
\phi_{S}(x_2)(\phi^S_{S}(x_3)-
 \phi^T_{S}(x_3))\right]E_{nfa}(t_f)h_{nfa}^f(x_1,x_2,x_3,b_1,b_2)
\right\} \label{eq:b-nfa-p1}\;,
\eeq

\item[]{(iii) $(S-P)(S+P)$ operators:}
\beq
M_{nfa}^{P2}&=& \frac{32}{\sqrt{6}}\pi C_F m_{B_{(s)}}^2
\int_{0}^{1}d x_{1}d x_{2}\,d x_{3}\,\int_{0}^{\infty} b_1d b_1 b_2d
b_2\,  
\non & &
\times \phi_{B_{(s)}}(x_1,b_1)\left\{
\left[(1-x_3)\phi_{S}(x_2)\phi_{S}(x_3)- r_S  r_S \left(\phi_{S}^S(x_2)
\right.\right.\right.
\non & &
\left.\left.\left. \times [(x_2-x_3+3)
\phi^S_{S}(x_3)-
(1-x_2-x_3)\phi^T_{S}(x_3)]+\phi_{S}^T(x_2)\right.\right.\right.
\non && \left.\left.\left.
\times [(1-x_2-x_3)\phi_{S}^S(x_3)+(1-x_2+x_3)\phi_{S}^T(x_3)]
\right)\right] E_{nfa}(t_f)  \right.\non
& & \left. \times
h_{nfa}^f(x_1,x_2,x_3,b_1,b_2)- E_{nfa}(t_e)h_{nfa}^e(x_1,x_2,x_3,b_1,b_2)
\right. \non && \left.
 \times
\left[x_2 \phi_{S}(x_2)\phi_{S}(x_3)+ r_S   r_S
\left(\phi_{S}^S(x_2)[(x_3-x_2-1) \phi^S_{S}(x_3)
\right.\right. \right.\non && \left. \left.\left.
-(1-x_2-x_3)\phi^T_{S}(x_3)]+\phi_{S}^T(x_2)
[(1-x_2-x_3)\phi_{S}^S(x_3)
\right.\right. \right.\non && \left. \left.\left.
+(1+x_2-x_3)\phi_{S}^T(x_3)]\right)\right]
 \right\} \label{eq:b-nfa-p2}\;;
\eeq
\end{enumerate}

For the last two diagrams in Fig.~\ref{fig:fig1}, i.e., the factorizable annihilation($fa$) diagrams
\ref{fig:fig1}(g) and \ref{fig:fig1}(h), we have
\begin{enumerate}
\item[]{(i) $(V-A)(V-A)$ operators:}
\beq
F_{fa}&=&  -8 \pi C_F m_{B_{(s)}}^2\int_0^1 d x_{2} dx_{3}\,
\int_{0}^{\infty} b_2 db_2 b_3 db_3\, \left\{ \left[ x_2
\phi_{S}(x_2) \phi_{S}(x_3)-2 r_S  r_S\right.\right. \non & &
\left.\left.  \times  \left((x_2 +
1)\phi^S_{S}(x_2)+(x_2-1)\phi^T_{S}(x_2)\right)\phi_{S}^S(x_3)\right]
 h_{fa}(x_2,1-x_3,b_2,b_3) \right. \non && \left.\times
E_{fa}(t_g)  - \left[(1-x_3)\phi_{S}(x_2) \phi_{S}(x_3)+2 r_S  r_S
\phi_{S}^S(x_2) \left((x_3-2)\phi^S_{S} (x_3)
\right.\right.\right.\non && \left.\left.\left.- x_3
\phi_{S}^T(x_3)\right) \right] E_{fa}(t_h)h_{fa}(1-x_3,x_2,b_3,b_2)
\right\}\label{eq:b-fa}\;,
\eeq

\item[]{(ii) $(V-A)(V+A)$ operators:}
\beq
 F_{fa}^{P1}&=& F_{fa}\;, \label{eq:b-fa-p1}
\eeq

\item[]{(iii) $(S-P)(S+P)$ operators:}
\beq
 F_{fa}^{P2}&=& -16 \pi C_F m_{B_{(s)}}^2  \int_0^1 d
x_{2} dx_{3}\, \int_{0}^{\infty} b_2 db_2 b_3 db_3\,\left\{ \left[2
r_S \phi_{S}(x_2) \phi^S_{S}(x_3) \right.\right. \non & &
\left.\left.-r_S
 x_2 (\phi_{S}^S(x_2)- \phi_{S}^T(x_2))\phi_{S}(x_3) \right]
h_{fa}(x_2,1-x_3,b_2,b_3) E_{fa}(t_g)\right. \non && \left.  + \left[ r_S
(1-x_3) \phi_{S}(x_2) (\phi_{S}^S(x_3)+\phi_{S}^T(x_3))- 2
r_S \phi_S^S(x_2)\phi_{S}(x_3) \right]
 \right. \non && \left. \times
E_{fa}(t_h)h_{fa}(1-x_3,x_2,b_3,b_2) \right\} \label{eq:b-fa-p2}\;.
 \eeq
\end{enumerate}
It is interesting to notice that there is a large cancelation in
the factorizable annihilation $F_{fa}$, i.e., Eq.~(\ref{eq:b-fa}), from the diagrams \ref{fig:fig1}(g)
and \ref{fig:fig1}(h), which can result in the tiny or small deviations from
zero and can be seen numerically as displayed in the last column of the 2nd line of Tables~\ref{tab:DA-s1} and
\ref{tab:DA-s2}. In the SU(3) limit, in particular, this cancelation will
lead to the exact zero contribution.

\subsection{$B_c \to {\ks}^+ {\ksb}^0$ decay }

In the SM, charmless hadronic $B_c \to {\ks}^+ {\ksb}^0$ decay can only occur through
the pure annihilation-type diagrams.
From the effective Hamiltonian~(\ref{eq:heff}), there are 4 types of
diagrams contributing to the $B_c \to {\ks}^+ {\ksb}^0$ decay as
illustrated in Fig.~\ref{fig:fig1}(e)-(h), which result in the Feynman
decay amplitudes $F'_{fa}$ and $M'_{nfa}$ from singly current-current operators,
respectively.
Following the same procedure as stated in the above subsection,
we can obtain the analytic decay amplitudes for $B_c \to {\ks}^+ {\ksb}^0$ mode,
\beq
 F'_{fa} &=&-8 \pi C_F m_{B_c}^2 \int_0^1 d x_{2} dx_{3}\,
 \int_{0}^{\infty} b_2 db_2 b_3 db_3\,
\non && \times
\left\{h_{fa}(1-x_{3},x_{2},b_{3},b_{2})E'_{fa}(t'_{g})
\left[x_{2} \phi_{S}(x_2)\phi_{S}(x_3)-2 r_S r_S\phi_{S}^S(x_3)
\right.\right. \non && \left.\left. \times
\left((x_2 + 1)\phi^{S}_{S}(x_2)+ (x_2 -1)\phi^{T}_{S}(x_{2})\right)
\right]- h_{fa}(x_{2},1-x_{3},b_{2},b_{3})E'_{fa}(t'_{h})
\right. \non && \left. \times
\left[ (1- x_3) \phi_{S}(x_2) \phi_{S}(x_3) + 2 r_S r_S \phi_{S}^S(x_2)
\left( (x_3 -2)\phi_{S}^S(x_3)- x_3 \phi_{S}^T(x_3)\right)\right]\right\}\;,
\label{eq:bc-fa}\\
 M'_{nfa} &=& -\frac{16 \sqrt{6}}{3}\pi C_F m_{B_c}^2 \int_{0}^{1}d x_{2}\,d x_{3}\,
 \int_{0}^{\infty} b_1d b_1 b_2d b_2\,
 \non && \times
 \left\{{h'}_{nfa}^{e}(x_2,x_3,b_1,b_2) E'_{nfa}(t'_e)
 \left[(r_c - x_3 +1) \phi_{S}(x_2)\phi_{S}(x_{3})- r_S r_S\left(\phi_{S}^S(x_2)
 \right.\right. \right.\non && \left. \left. \left. \times
 ((3 r_c + x_2 -x_3 +1) \phi_{S}^S(x_3)-(r_c -x_2 -x_3 +1)\phi_{S}^T(x_3))+\phi_{S}^T(x_2)
 \right.\right.\right. \non && \left.\left. \left. \times
 ((r_c-x_2 -x_3 +1) \phi_{S}^S(x_3)+(r_c -x_2 +x_3 -1)\phi_{S}^T(x_3))\right)\right]-E'_{nfa}(t'_f)
 \right. \non && \left. \times
 \left[ (r_b + r_c +x_2 -1) \phi_{S}(x_2) \phi_{S}(x_3) - r_S r_S \left(\phi_{S}^S(x_2)((4 r_b +r_c +x_2 -x_3
 \right.\right.\right.\non && \left. \left.\left.
 -1)\phi_{S}^S(x_3)-(r_c + x_2 +x_3 -1)\phi_{S}^T(x_3)) +\phi_{S}^T(x_2)((r_c + x_2 +x_3 -1)
 \right.\right.\right. \non && \left.\left.\left. \times
 \phi_{S}^S(x_3)-(r_c +x_2 -x_3 -1) \phi_{S}^T(x_3))\right)\right] {h'}_{nfa}^{f}(x_2,x_3,b_1,b_2)\right\}\;,
 \label{eq:bc-nfa}
 \eeq
where the non-relativistic approximation form of the distribution amplitude $\phi_{B_c}$ for $B_c$ meson
has been used, the convolution factor $E'_i$, the hard scale $t'_i$, and the hard function $h'_i$
are referred to Refs.~\cite{Liu10:bc2s,Xiao:pqcd1}. Moreover, $r_{b}= m_{b}/m_{B_c}$,
$r_{c}= m_{c}/m_{B_c}$, and $r_b+r_c \approx 1$ in $B_c$ meson.

\subsection{Decay Amplitudes for $B_q \to \ks \ksb (q=u, d, s, c)$ Channels}

By combining various of contributions from the relevant Feynman
diagrams together, the total decay amplitudes for the four
penguin-dominated decays $B_{(s)} \to \ks \ksb$ and the pure annihilation processes $B_d \to {\ks}^+ {\ks}^-$ and $B_c \to {\ks}^+ {\ksb}^0$ can then read as,
\begin{enumerate}

\item {The total decay amplitudes of $B \to \ks \ksb$ decays:}
\beq
{\cal A}(B_u \to {\ks}^+ {\ksb}^0) &=& \lambda_u \biggl[ 
M_{nfa} C_1\biggr]
- \lambda_t \biggl[F_{fs} (a_4- \frac{1}{2} a_{10})
+ F_{fs}^{P2} (a_6 -\frac{1}{2} a_8)  \non
&&  + M_{nfs} (C_3- \frac{1}{2} C_9) + M_{nfs} (C_5 -\frac{1}{2} C_7)
+ M_{nfa}  \non &&
\times  (C_3 + C_9)+ M_{nfa}^{P1} (C_5 + C_7) 
+ f_B F_{fa}^{P2} (a_6 +a_8)\biggr]\;, \label{eq:tda-b2kpk0b}
\eeq
where $\lambda_{u}= V_{ub}^* V_{ud}$ and $\lambda_t = V_{tb}^* V_{td}$.
\beq
{\cal A}(B_d \to {\ks}^+ {\ks}^-) &=& \lambda_u \biggl[ M_{nfa} C_2\biggr]-\lambda_t \biggl[
 M_{nfa} (C_4 + C_{10}) + M_{nfa}^{P2} (C_6 + C_8) \non &&
 + M_{nfa}[{\ks}^+ \leftrightarrow {\ks}^-]
 (C_4 - \frac{1}{2} C_{10})
\non &&
 + M_{nfa}^{P2}[{\ks}^+ \leftrightarrow {\ks}^-] (C_6 - \frac{1}{2} C_8) \biggr]\;,\label{eq:tda-b2kpkm}
\eeq
\beq
{\cal A}(B_d \to {\ks}^0 {\ksb}^0) &=& -\lambda_t \biggl[ F_{fs}
(a_4 - \frac{1}{2} a_{10}) + F_{fs}^{P2} (a_6 - \frac{1}{2} a_8)
+ (M_{nfs} + M_{nfa} ) \non &&
\times (C_3 -\frac{1}{2} C_9)+(M_{nfs}^{P1}  + M_{nfa}^{P1})(C_5 - \frac{1}{2} C_7)
+ (M_{nfa} \non &&
 + [{\ks}^0 \leftrightarrow {\ksb}^0] ) (C_4 - \frac{1}{2} C_{10})
 + (M_{nfa}^{P2} + [{\ks}^0 \leftrightarrow {\ksb}^0] ) \non &&
\times (C_6 - \frac{1}{2} C_8)
 + f_B F_{fa}^{P2} (a_6 - \frac{1}{2}a_8) \biggr]\;;\label{eq:tda-b2k0k0b}
\eeq

\item {The total decay amplitudes of $B_s \to \ks \ksb$ decays:}
\beq
{\cal A}(B_s \to {\ks}^+ {\ks}^-) &=&
\lambda'_u \biggl[ F_{fs} a_1 + M_{nfs} C_1 + M_{nfa} C_2\biggr]
-\lambda'_t \biggl[ F_{fs} (a_4 + a_{10})  \non &&
+ F_{fs}^{P2} (a_6 +a_8)
+ M_{nfs} (C_3 + C_9) + M_{nfs}^{P1} (C_5 + C_7)
  \non &&
 +  M_{nfa}(C_3 - \frac{1}{2}C_9 + C_4 - \frac{1}{2} C_{10})
+ M_{nfa}[{\ks}^+ \leftrightarrow {\ks}^-] \non &&
\times (C_4 + C_{10})
+ M_{nfa}^{P1} (C_5 - \frac{1}{2} C_7)
+ M_{nfa}^{P2} (C_6 - \frac{1}{2} C_8)\non &&
+ M_{nfa}^{P2}[{\ks}^+ \leftrightarrow {\ks}^-] (C_6 + C_8)
+ f_{B_s} F_{fa}^{P2} (a_6 - \frac{1}{2} a_8) \biggr]\;,\label{eq:tda-bs2kpkm}
\eeq
where $\lambda'_{u}= V_{ub}^* V_{us}$ and $\lambda'_t = V_{tb}^* V_{ts}$.
\beq
{\cal A}(B_s \to {\ks}^0 {\ksb}^0) &=& -\lambda'_t \biggl[ F_{fs}
(a_4 - \frac{1}{2} a_{10}) + F_{fs}^{P2} (a_6 - \frac{1}{2} a_8)
+ (M_{nfs} + M_{nfa} )\non &&
\times (C_3 -\frac{1}{2} C_9)
+(M_{nfs}^{P1}  + M_{nfa}^{P1})(C_5 - \frac{1}{2} C_7)
 + (M_{nfa} \non &&
 + [{\ks}^0 \leftrightarrow {\ksb}^0] )
 (C_4 - \frac{1}{2} C_{10})
 + (M_{nfa}^{P2} + [{\ks}^0 \leftrightarrow {\ksb}^0] )\non &&
 \times (C_6 - \frac{1}{2} C_8)
 +f_{B_s} F_{fa}^{P2} (a_6 -\frac{1}{2} a_8) \biggr]\;;\label{eq:tda-bs2k0k0b}
\eeq

\item {The total decay amplitude of $B_c \to
{\ks}^+ {\ksb}^0$ decay:}
\beq {\cal A}(B_c \to {\ks}^+ {\ksb}^0)
&=& V_{cb}^* V_{ud} \biggl[f_{B_c} F'_{fa} a_1 + M'_{nfa}
C_1\biggr] \;.\label{eq:tda-bc2kpk0b}
\eeq
\end{enumerate}
In the above decay amplitudes for the channels $B_{(s)} \to \ks \ksb$, based on the relevant discussions after Eq.~(\ref{eq:b-fa-p2}), we have neglected the factorizable annihilation contributions $F_{fa}$ in Eqs.~(\ref{eq:tda-b2kpk0b})-(\ref{eq:tda-bs2k0k0b})
 induced from the small SU(3) symmetry breaking effects.
However, for the
pure annihilation mode $B_c \to {\ks}^+ {\ksb}^0$, to present the large annihilation
contribution occurred in this considered $B_c$ decay channel, we therefore have
kept the factorizable decay amplitude $F'_{fa}$
in Eq.~(\ref{eq:tda-bc2kpk0b}). 

\section{Numerical Results and Discussions}\label{sec:3}

In this section, we will make the theoretical predictions on the
CP-averaged BRs and CP-violating asymmetries for those
considered $B_q \to \ks \ksb$ decay modes.
 In numerical calculations, central values of the input parameters will be
used implicitly unless otherwise stated. Firstly, we shall make several essential
discussions on the input quantities.

\subsection{Input Quantities}
The pQCD predictions depend on the inputs for the nonperturbative parameters
such as decay constants and universal distribution amplitudes for heavy
pseudoscalar $B_q$ and light scalar $\ks$ mesons.
For the $B/B_s$ mesons, the distribution amplitudes in the $b$
space have been proposed
\beq
\phi_{B}(x,b)&=& N_Bx^2(1-x)^2
\exp\left[-\frac{1}{2}\left(\frac{xm_B}{\omega_b}\right)^2
-\frac{\omega_b^2 b^2}{2}\right] \;,
\eeq
in Refs.~\cite{Li01:kpi,Lu01:pipi} and
\beq
\phi_{B_s}(x,b)&=& N_{B_s} x^2(1-x)^2
\exp\left[-\frac{1}{2}\left(\frac{xm_{B_s}}{\omega_{bs}}\right)^2
-\frac{\omega_{bs}^2 b^2}{2}\right] \;,
\eeq
in Ref.~\cite{Ali07:bsnd}, respectively, where the normalization factors $N_{B_{(s)}}$
are related to the decay constants $f_{B_{(s)}}$ through
\beq
\int_0^1 dx \phi_{B_{(s)}}(x, b=0) &=& \frac{f_{B_{(s)}}}{2 \sqrt{6}}\;.
\eeq
In recent years, a lot of studies for $B$ decays have been performed
in the pQCD approach~\cite{Li01:kpi,Lu01:pipi}, and the shape parameter
$\omega_b$ has been fixed at $0.40$~GeV by using the rich experimental
data on the $B$ mesons with $f_{B}= 0.19$~GeV. Correspondingly, the normalization constant $N_B$
is $91.745$. For $B_s$ meson, considering a small SU(3) symmetry breaking,
since $s$ quark is heavier than the $u$ or $d$ quark,
the momentum fraction of $s$ quark should be a little larger than that
of $u$ or $d$ quark in the $B$ mesons, we therefore adopt the shape parameter
$\omega_{bs} = 0.50$~GeV~\cite{Ali07:bsnd} with $f_{B_s} = 0.23$~GeV, then the corresponding normalization
constant is $N_{B_s} = 63.67$. In order to analyze the uncertainties of
theoretical predictions induced
by the inputs, we can vary the shape parameters $\omega_{b}$ and $\omega_{bs}$ by
10\%, i.e., $\omega_b = 0.40 \pm 0.04$~GeV and $\omega_{bs} = 0.50 \pm 0.05$~GeV, respectively.

As for the double heavy-flavored $B_c$ meson, since it embraces two heavy quarks $b$ and $c$ simultaneously,
the distribution amplitude $\phi_{B_c}$ would be close to
$\delta(x- m_c/m_{B_c})$~\cite{Bell08:bcda} in the
non-relativistic limit, we therefore adopt the non-relativistic
approximation form(See~\cite{Xiao:pqcd1}
and references therein),
\beq
\phi_{B_c}(x) &=& \frac{f_{B_c}}{2 \sqrt{6}} \delta(x - m_c/ m_{B_c})\;,
\eeq
where $f_{B_c}$ is the decay constant for $B_c$ meson and to be determined by
the precision data in principle. Unfortunately, however, there are no available
experimental measurements on $f_{B_c}$, we then adopt the result calculated in
 Lattice QCD~\cite{Chiu07:fbc},
\beq
f_{B_c} &=& (489 \pm 4 \pm 3) ~~{\rm MeV}\;.
\eeq

For the scalar $\ks$, its leading twist light-cone distribution amplitude
$\phi_{\ks}(x,\mu)$ can be generally expanded as the Gegenbauer
polynomials~\cite{Cheng06:B2SP,Li09:B2Sfm}:
\beq
\phi_{\ks}(x,\mu)&=&\frac{3}{\sqrt{2N_c}}x(1-x)\biggl\{f_{\ks}(\mu)+\bar
f_{\ks}(\mu)\sum_{m=1}^\infty B_m(\mu)C^{3/2}_m(2x-1)\biggr\}, \eeq
where $f_{\ks}(\mu)$ and $\bar f_{\ks}(\mu)$, $B_m(\mu)$, and
$C_m^{3/2}(t)$ are the vector and scalar decay constants,
Gegenbauer moments, and Gegenbauer polynomials,
respectively.

There exists a relation between
the vector and scalar decay constants,
\beq
 \bar f_{\ks} &=& \mu_{\ks} f_{\ks} \;\;\;\;  {\rm and} \;\;\;\;
 \mu_{\ks} = \frac{m_{\ks}}{m_2(\mu)-m_1(\mu)}\;,
\eeq
where $m_1$ and $m_2$ are the
running current quark masses in the scalar $\ks$.

The values for scalar decay constants and Gegenbauer moments in
the distribution amplitudes of $\ks$ have been investigated at scale $\mu=1~
\mbox{GeV}$ in scenarios S1 and S2~\cite{Cheng06:B2SP}:
\beq
\bar f_{K_0^*}&=& -0.300 \pm 0.030~{\rm GeV}, \quad B_1=\;\;\; 0.58 \pm 0.07, \quad  B_3= -1.20 \pm 0.08\;\;\; \rm{(S1)}\;,  \non
\bar f_{K_0^*}&=&\;\;\; 0.445 \pm 0.050~{\rm GeV}, \quad B_1= -0.57 \pm 0.13, \quad  B_3= -0.42 \pm 0.22\;\;\; \rm{(S2)}\;;
\eeq

As for the twist-3 distribution amplitudes $\phi_{\ks}^S$ and
$\phi_{\ks}^T$, they have been investigated in only S2 with large
uncertainties~\cite{Lu07:3DAs-scalar}. We therefore adopt the asymptotic forms
in our numerical calculations for simplicity:
\beq
\phi^S_{\ks}&=& \frac{1}{2\sqrt {2N_c}}\bar f_{\ks},\,\,\,\,\,\,\,\qquad
\phi_{\ks}^T=
\frac{1}{2\sqrt {2N_c}}\bar f_{\ks}(1-2x).
\eeq

The QCD scale~({\rm GeV}), masses~({\rm GeV}), 
 and $B_q$ meson lifetime({\rm ps}) are~\cite{Li01:kpi,Lu01:pipi,Amsler08:pdg}
\beq
 \Lambda_{\overline{\mathrm{MS}}}^{(f=4)} &=& 0.250\; , \qquad m_W = 80.41\;,
 \qquad m_{B_c} = 6.286\;,  \qquad  m_{B_s}= 5.366\;; \non
   m_B &=& 5.279\;,\hspace{0.62cm} \quad m_b = 4.8\;,\hspace{0.67cm} \qquad m_c = 1.5\;,
   \qquad \hspace{0.33cm}  m_{\ks} =1.425\;;  \non
  \tau_{B_u} &=& 1.638\;, \hspace{0.13cm}
\qquad \tau_{B_d}= 1.53\;,\qquad \hspace{0.43cm} \tau_{B_s} = 1.47\;,\qquad \hspace{0.36cm}
\tau_{B_c}= 0.46\;. \label{eq:mass}
\eeq

For the CKM matrix elements, we adopt the Wolfenstein
parametrization and the updated parameters $A=0.814$,
 $\lambda=0.2257$, $\bar{\rho}=0.135$, and $\bar{\eta}=0.349$~\cite{Amsler08:pdg}.

Utilizing the above chosen distribution amplitudes and the central values of the relevant
input parameters, we can get the numerical results in the pQCD approach for the form
factors $F_{0,1}^{B \to \ks}$
and $F_{0,1}^{B_s \to \ksb}$ from Eq.~(\ref{eq:b-fs}) at maximal recoil
 as follows,
  \beq
F_{0,1}^{B \to \ks} (q^2=0) &=&  \left\{ \begin{array}{ll}
-0.34^{+0.04}_{-0.06}(\omega_{b})^{+0.03}_{-0.04}(\bar f_S)^{+0.01}_{-0.03}(B_{i}^{S}) & \quad ({\rm S1}) \\
\;\;\; 0.63^{+0.10}_{-0.08}(\omega_{b})^{+0.07}_{-0.07}(\bar f_S)^{+0.06}_{-0.06}(B_{i}^{S})  & \quad ({\rm S2}) \\ \end{array} \right.   \label{eq:formf-b}  \;,\\
F_{0,1}^{B_s \to \ksb}(q^2=0) &=&  \left\{ \begin{array}{ll}
-0.31^{+0.05}_{-0.05}(\omega_{bs})^{+0.03}_{-0.03}(\bar f_S)^{+0.01}_{-0.01}(B_{i}^{S}) & \quad ({\rm S1}) \\
\;\;\; 0.57^{+0.09}_{-0.08}(\omega_{bs})^{+0.06}_{-0.07}(\bar f_S)^{+0.05}_{-0.06}(B_{i}^{S}) & \quad ({\rm S2}) \\ \end{array} \right.   \label{eq:formf-bs} \;.
  \eeq
where the errors arise from the shape parameter $\omega_b(\omega_{bs})$ in $B(B_s)$
meson distribution amplitude, the scalar decay constant $\bar f_S$, and the Gegenbauer moments
$B_i^S$ in the light $\ks$ distribution amplitude, respectively.
These values agree well with those as given in Ref.~\cite{Li09:B2Sfm}.

\subsection{ CP-averaged Branching Ratios }

In this subsection, we will analyze the CP-averaged BRs
of the cahrmless hadronic $B_q \to \ks \ksb$ decays in
the pQCD approach.
For $B_q \to \ks \ksb$ decays, the decay rate can be written as
\beq
\Gamma =\frac{G_{F}^{2}m^{3}_{B_q}}{32 \pi  } (1-2 r_S^2) |{\cal A}(B_q
\to \ks \ksb)|^2\;,\label{eq:bqdr}
\eeq
where the corresponding decay amplitudes ${\cal A}$ have been
given explicitly in Eqs.~(\ref{eq:tda-b2kpk0b}-\ref{eq:tda-bc2kpk0b}).
Using the decay amplitudes
obtained in last section, it is straightforward to calculate the
CP-averaged BRs with uncertainties as displayed in
Eqs.~(\ref{eq:bru})-(\ref{eq:brd2}), (\ref{eq:brs1})-(\ref{eq:brs2}), and (\ref{eq:bcss}).
The dominant errors are induced by the uncertainties of the shape parameters
$\omega_b = 0.40 \pm 0.04$~GeV for $B$ mesons and $\omega_{bs} = 0.50 \pm 0.05$~GeV for $B_s$ meson,
the charm quark mass $m_c=1.5 \pm 0.15$~GeV for $B_c$ meson, the scalar decay constants $\bar f_S$ and
the Gegenbauer moments $B^S_i(i=1, 3)$ for the scalar $\ks$, 
and CKM matrix elements ($\bar \rho, \bar \eta$), respectively.
It is worth of mentioning that the variation of the CKM parameters has a little effects
on the BRs of these considered $B_q \to \ks \ksb$ decays
in the pQCD approach and thus will be neglected in the numerical results as shown in
Eqs.~(\ref{eq:bru})-(\ref{eq:bcss}).

The pQCD predictions for the CP-averaged BRs of the
decays under consideration within errors in both scenarios S1 and S2 are the following,
\beq
Br(B_u \to {\ks}^+ {\ksb}^0) &\approx& \left\{ \begin{array}{ll}
1.1^{+0.1}_{-0.2}(\omega_{b})^{+0.5}_{-0.4}(\bar f_S)^{+0.1}_{-0.2}(B_{i}^{S})
\times  10^{-5}& \quad ({\rm S1}) \\
1.9^{+0.2}_{-0.1}(\omega_{b})^{+1.1}_{-0.7}(\bar f_S)^{+1.3}_{-0.9}(B_{i}^{S})
\times  10^{-5}& \quad ({\rm S2}) \\ \end{array} \right.
\label{eq:bru},
 \eeq
 \beq
Br(B_d \to {\ks}^0 {\ksb}^0) &\approx& \left\{
\begin{array}{ll} 1.1^{+0.2}_{-0.2}(\omega_{b})^{+0.6}_{-0.4}(\bar
f_S)^{+0.2}_{-0.1}(B_{i}^{S})
\times  10^{-5}& \quad ({\rm S1}) \\
2.0^{+0.0}_{-0.1}(\omega_{b})^{+1.0}_{-0.8}(\bar f_S)^{+1.7}_{-1.2}(B_{i}^{S})
\times  10^{-5}& \quad ({\rm S2}) \\ \end{array} \right.   \label{eq:brd1},\\
Br(B_d \to {\ks}^+ {\ks}^-) &\approx&  \left\{ \begin{array}{ll}
3.2^{+0.0}_{-0.1}(\omega_{b})^{+1.6}_{-1.1}(\bar f_S)^{+1.0}_{-0.8}(B_{i}^{S})
\times  10^{-6}& \quad ({\rm S1}) \\
2.2^{+0.4}_{-0.4}(\omega_{b})^{+1.1}_{-0.8}(\bar f_S)^{+1.8}_{-1.2}(B_{i}^{S})
\times  10^{-6}& \quad ({\rm S2}) \\ \end{array} \right.   \label{eq:brd2};
\eeq
\beq
Br(B_s \to {\ks}^0 {\ksb}^0) &\approx&  \left\{ \begin{array}{ll}
2.5^{+0.6}_{-0.5}(\omega_{bs})^{+1.1}_{-0.9}(\bar f_S)^{+0.4}_{-0.4}(B_{i}^{S})
\times  10^{-4}& \quad ({\rm S1}) \\
5.4^{+0.1}_{-0.2}(\omega_{bs})^{+2.8}_{-2.1}(\bar f_S)^{+4.8}_{-3.3}(B_{i}^{S})
\times  10^{-4}& \quad ({\rm S2}) \\ \end{array} \right.   \label{eq:brs1}  ,\\
Br(B_s \to {\ks}^+ {\ks}^-) &\approx& \left\{ \begin{array}{ll}
2.3^{+0.5}_{-0.4}(\omega_{bs})^{+1.1}_{-0.8}(\bar f_S)^{+0.4}_{-0.4}(B_{i}^{S})
\times  10^{-4}& \quad ({\rm S1}) \\
5.2^{+0.2}_{-0.2}(\omega_{bs})^{+2.8}_{-2.0}(\bar f_S)^{+4.6}_{-3.1}(B_{i}^{S})
\times  10^{-4}& \quad ({\rm S2}) \\ \end{array} \right.   \label{eq:brs2}  ;
\eeq
\beq
Br(B_c \to {\ks}^+ {\ksb}^0) &\approx& \left\{ \begin{array}{ll}
2.1^{+0.7}_{-0.6}(m_c)^{+1.0}_{-0.7}(\bar f_S)^{+0.5}_{-0.4}(B_{i}^{S})
 \times  10^{-4}& \quad ({\rm S1}) \\
3.0^{+0.7}_{-0.6}(m_c)^{+1.6}_{-1.1}(\bar f_S)^{+3.2}_{-1.8}(B_{i}^{S})
 \times  10^{-5}& \quad ({\rm S2}) \\ \end{array} \right.   \label{eq:bcss}  \;.
\eeq
It is easy to see that above BRs are rather large, in the range of $10^{-6}$ to $10^{-4}$,
and can be  measured at B factories for the relevant $B_{u,d}$ decays,
and at the LHC experiments for  the considered $B_s/B_c$ decay modes.
More importantly, the pure annihilation decays $B_d \to {\ks}^+ {\ks}^-$ and
$B_c \to {\ks}^+ {\ksb}^0$ with large BRs can provide important information about
the weak annihilation amplitudes and more evidences on the sizable annihilation
contributions in B physics,
then further shed light on the corresponding annihilation decay mechanism.

By comparison with those of $B/B_s  \to K\overline{K}$
decays~\cite{Ali07:bsnd,Chen00:b2kk,Beneke03:bdecay,Cheng09:b2m2m3,Cheng09:bs2m2m3},
one can easily find that the pQCD predictions for the BRs of
$B/B_s \to \ks \ksb$ channels are much larger than that for the
corresponding $B/B_s \to K \overline{K}$ decays by generally a factor of 10.
For the pure annihilation mode $B_d \to {\ks}^+ {\ks}^-$, however, the enhancement factor
is about 100.
The main reason for so large difference is that the QCD behavior for the $p$-wave
scalar $\ks$ is very different from that for the $s$-wave pseudoscalar kaon,
which can be seen clearly from their distribution amplitudes~\cite{Cheng06:B2SP,Pseudoscalar}:
the former governed by the odd Gegenbauer moments, while the latter dominated by
the even ones, apart from the small symmetry breaking term $a_1^K$.

As mentioned in the introduction, up to now, there is
only one preliminary upper limit on the branching ratio of $B^0 \to {\ks}^0 {\ksb}^0$ at
the 90\% confidence level ~\cite{Chiang10:k0k0b,Chen11:belle,Asner10:hfag}:
\beq Br(B^0 \to {\ks}^0 {\ksb}^0) &<&
8.4 \times 10^{-6}\;,\label{eq:brks0ks0-ex}
\eeq
While in the pQCD approach, our theoretical predictions of the BRs for $B^0 \to
{\ks}^0 {\ksb}^0$ decay in both scenarios within errors are as
follows,
\beq Br(B^0 \to {\ks}^0 {\ksb}^0) &\approx& \left\{
\begin{array}{ll}
6 \sim 18\;\;  \times  10^{-6}& \quad ({\rm S1}) \\
6 \sim 40\;\;  \times  10^{-6}& \quad ({\rm S2}) \\ \end{array}
\right.   \label{eq:brks0ks0-th},
\eeq
It is easy to see that the pQCD predictions within theoretical errors in both scenarios
are consistent with currently available experiment upper limit, and will be tested directly
when more data samples are collected at the LHC experiments.

As shown in Eqs.~(\ref{eq:bru})-(\ref{eq:bcss}), the pQCD predictions have a strong dependence
on the input parameters describing the nonperturbative behavior of
the light scalar $\ks$ ( the scalar decay constant $\bar f_S$ and Gegenbauer
moments $B_i^S$), and a moderate dependence on the shape parameters
$\omega_{b}(\omega_{bs})$ in $\phi_B(\phi_{B_s})$ and the charm quark mass $m_c$ in $\phi_{B_c}$.
Therefore, once the relevant experiments
could provide the precise measurements for $B \to \ks \ksb$ modes,
one can has a better understanding about the nonperturbative hadron dynamics in $\ks$.

In order to facilitate the
discussion, we present the central values of our predictions
in terms of the topological amplitudes in Tables~\ref{tab:DA-s1} and \ref{tab:DA-s2},
where ${\cal F}_{fs}$, ${\cal F}_{fa}$, ${\cal M}_{nfs}$ and ${\cal M}_{nfa}$ denote
the decay amplitudes
from factorizable spectator, factorizable annihilation, nonfactorizable spectator,
and nonfactorizable annihilation contributions, respectively.

\begin{table}[ht]
\caption{ The factorization decay amplitudes(in unit of $10^{-3}\; \rm{GeV}^3$) of the
charmless hadronic $B_q \to \ks \ksb$
decays 
in S1, where only the central values are quoted for clarification.}
\label{tab:DA-s1}
\begin{center}\vspace{-0.6cm}
{\small \begin{tabular}[t]{l|c|c|c|c}
 \hline \hline
 Channels     &    ${\cal F}_{fs}$      &  ${\cal M}_{nfs}$          &  ${\cal M}_{nfa}^{(')}$              &    ${\cal F}^{(')}_{fa}$           \\
\hline \hline
$B_u \to {\ks}^+ {\ksb}^0$              &$ 1.246- {\it i} 0.518$   &$ 0.468- {\it i} 0.729$  &$ -0.733+ {\it i} 0.335$  &$ 2.540+ {\it i} 1.614$
\\
$B_d \to {\ks}^+ {\ks}^-$             &$0.0$  &$ 0.0$  &$ 2.070- {\it i} 2.517$  &$ -0.013+ {\it i} 0.009$
\\
$B_d \to {\ks}^0 {\ksb}^0$             &$1.246- {\it i} 0.518$  &$ 0.468- {\it i} 0.729$  &$ 0.269- {\it i} 1.118$  &$ 2.690+ {\it i} 1.741$
 \\
  \hline \hline
$B_s \to {\ks}^+ {\ks}^-$         &$-6.020+ {\it i} 0.190$  &$ -0.736+ {\it i} 1.594$  &$ -2.267+ {\it i} 5.741$  &$ -10.190- {\it i} 14.192$
 \\
$B_s \to {\ks}^0 {\ksb}^0$       &$-6.270$  &$ -2.732+ {\it i} 2.310$  &$ -2.727+ {\it i} 6.376$  &$ -10.187- {\it i} 14.273$
\\
 \hline \hline
$B_c \to {\ks}^+ {\ksb}^0$               &$0.0$  &$ 0.0$  &$ 35.744- {\it i} 11.105$  &$ 1.658- {\it i} 2.530$
 \\
\hline \hline
\end{tabular} }
\end{center}
\end{table}
\begin{table}[ht]
\caption{ Same as Table~\ref{tab:DA-s1} but in S2.}
\label{tab:DA-s2}
\begin{center}\vspace{-0.6cm}
{\small \begin{tabular}[t]{l|c|c|c|c}
 \hline \hline
 Channels     &    ${\cal F}_{fs}$      &  ${\cal M}_{nfs}$          &  ${\cal M}_{nfa}^{(')}$              &    ${\cal F}^{(')}_{fa}$           \\
\hline \hline
$B_u \to {\ks}^+ {\ksb}^0$              &$ 2.759- {\it i} 1.147$   &$ -0.364+ {\it i} 0.377$  &$ 0.601+ {\it i} 0.407$  &$ 2.528+ {\it i} 4.018$
\\
$B_d \to {\ks}^+ {\ks}^-$             &$0.0$  &$ 0.0$  &$ -1.934+ {\it i} 0.527$  &$ 0.004- {\it i} 0.001$
\\
$B_d \to {\ks}^0 {\ksb}^0$             &$ 2.759- {\it i} 1.147$  &$ -0.364+ {\it i} 0.377$  &$ -0.800+ {\it i} 1.260$  &$ 2.629+ {\it i} 4.035$
 \\
  \hline \hline
$B_s \to {\ks}^+ {\ks}^-$         &$-13.606+ {\it i} 0.506$  &$ 0.833- {\it i} 1.071$  &$ 4.562- {\it i} 4.863$  &$ -5.022- {\it i} 25.089$
 \\
$B_s \to {\ks}^0 {\ksb}^0$       &$-14.158$  &$ 2.289- {\it i} 0.961$  &$ 4.968- {\it i} 4.847$  &$ -5.293- {\it i} 25.192$
\\
 \hline \hline
$B_c \to {\ks}^+ {\ksb}^0$               &$0.0$  &$ 0.0$  &$ -6.626- {\it i} 14.063$  &$ 1.387+ {\it i} 0.083$
 \\
\hline \hline
\end{tabular}}
\end{center}
\end{table}

Based on the numerical results as shown in Tables~\ref{tab:DA-s1} and \ref{tab:DA-s2},
one can easily observe that
the factorizable contributions ${\cal F}_{fs}$ and ${\cal F}_{fa}$
govern the considered four penguin-dominated
decay modes $B_u \to {\ks}^+ {\ksb}^0$, $B_d \to {\ks}^0 {\ksb}^0$, $B_s \to {\ks}^+ {\ks}^-$,
and $B_s \to {\ks}^0 {\ksb}^0$, where both ${\cal F}_{fs}$\footnote{For the considered decays,
$F_{fs}^{P2}$ with scalar decay constant $\bar f_S$ contributes to
 the magnitude of ${\cal F}_{fs}$ dominantly. This is just because the contributions arising
 from the vector current operators are very small with the strongly suppressed vector decay
 constant $f_S$.}
and ${\cal F}_{fa}$ are mainly determined by the
contributions induced by $(S-P)(S+P)$ operators, e.g., see
Eqs.~(\ref{eq:tda-b2kpk0b}) and (\ref{eq:tda-b2k0k0b}-\ref{eq:tda-bs2k0k0b});
while the nonfactorizable contributions ${\cal M}^{(')}_{nfa}$
dominate the rest pure annihilation channel $B_d \to {\ks}^+ {\ks}^-(B_c \to {\ks}^+ {\ksb}^0)$.
The small magnitudes of the factorizable annihilation contributions in
$B_d \to {\ks}^+ {\ks}^-$ and $B_c \to {\ks}^+ {\ksb}^0$ decays are induced by
the tiny SU(3) symmetry breaking effects.
The difference between ${\cal F}_{fa}(B_d \to {\ks}^+ {\ks}^-)$ and
${\cal F}'_{fa}(B_c \to {\ks}^+ {\ksb}^0)$ in two scenarios
as presented in Tables~\ref{tab:DA-s1} and \ref{tab:DA-s2}, however,
is mainly governed by two factors apart from $f_{B_c}/f_{B} \approx 2.6$:
one is the ratio of $V_{cb}/V_{ub} \approx 11.5$,
the other is the large ratio of $a_1/a_2 \approx 1000$ at the characteristic hard
scale $m_b/2$~\cite{Ali07:bsnd}.
The above phenomenological analysis can also be applied to the nonfactorizable annihilation
topology.
Anyway, one can see that these two considered pure annihilation modes are determined by
the contributions arising from the nonfactorizable diagrams, i.e.,
${\cal M}_{nfa}$ and ${\cal M}'_{nfa}$ (See Tables~\ref{tab:DA-s1} and \ref{tab:DA-s2}),
respectively.

It should be stressed that all these considered $B_q \to \ks \ksb$ decays have
large weak annihilation contributions, which can be seen clearly in
Tables~\ref{tab:DA-s1} and \ref{tab:DA-s2}. It is expected that these decay channels
with precision measurements at the experiments
will offer a good platform to study the underlying annihilation mechanism.

Using the pQCD predictions, we get the following interesting information on
the ratio of the BR's for two sets of considered decays in both scenarios S1 and S2,
\beq
\frac{\tau_{B_d}}{\tau_{B_u}} \cdot
\frac{Br(B_u \to {\ks}^+ {\ksb}^0)_{\rm pQCD}}{Br(B_d \to {\ks}^0 {\ksb}^0)_{\rm pQCD}}
& \approx & 0.9 \;,
\qquad \frac{Br(B_s \to {\ks}^0 {\ksb}^0)_{\rm pQCD}}{Br(B_s \to {\ks}^+ {\ks}^-)_{\rm pQCD}}
\approx 1\;.
\eeq
If these ratios are measured in the near future,
they can offer a great opportunity to study the QCD dynamics involved in these
four decay channels and will be helpful to test the adopted scenario of the scalar
$\ks$ in this work.

On the other hand, it is very interesting to notice the ratios among
the theoretical BRs of the considered modes in the pQCD approach
 \beq
 \frac{Br(B_u \to {\ks}^+ {\ksb}^0)_{\rm S2}}{Br(B_u \to {\ks}^+ {\ksb}^0)_{\rm S1}} &\approx&
\frac{Br(B_d \to {\ks}^0 {\ksb}^0)_{\rm S2}}{Br(B_d \to {\ks}^0 {\ksb}^0)_{\rm S1}} \approx (1.7 \sim 1.8)\;,
\label{eq:br-re-b}
 \eeq
  \beq
 \frac{Br(B_s \to {\ks}^+ {\ks}^-)_{\rm S2}}{Br(B_s \to {\ks}^+ {\ks}^-)_{\rm S1}} &\approx&
\frac{Br(B_s \to {\ks}^0 {\ksb}^0)_{\rm S2}}{Br(B_s \to {\ks}^0 {\ksb}^0)_{\rm S1}} \approx (2.2 \sim 2.3)\;,
\label{eq:br-re-bs}
 \eeq
 \beq
 \frac{Br(B_d \to {\ks}^+ {\ks}^-)_{\rm S1}}{Br(B_d \to {\ks}^+ {\ks}^-)_{\rm S2}} &\approx&
 1.5 \;,  \qquad
 \frac{Br(B_c \to {\ks}^+ {\ksb}^0)_{\rm S1}}{Br(B_c \to {\ks}^+ {\ksb}^0)_{\rm S2}} \approx
 7.0 \;. \label{eq:br-re-bca}
 \eeq
where the central values for the CP-averaged BRs have been quoted.
As given in Eqs.~(\ref{eq:br-re-b})-(\ref{eq:br-re-bca}), one can easily find that the CP-averaged BRs for the weak annihilation processes $B_d \to {\ks}^+ {\ks}^-$ and $B_c \to {\ks}^+ {\ksb}^0$ in S1 are larger than those in S2 to different extent. While for other four penguin-dominated $B_u \to {\ks}^+ {\ksb}^0$, $B_d \to {\ks}^0 {\ksb}^0$, $B_s \to {\ks}^0 {\ksb}^0$, and $B_s \to {\ks}^+ {\ks}^-$ channels, the CP-averaged BRs in the second scenario are larger than those in the first one with a factor around 2. These two different patterns might reveal the different QCD dynamics involved in the corresponding decay channels. The above relevant relations can be confronted with the near future experiments.

\subsection{CP-violating Asymmetries }

Now we turn to the evaluations of the CP-violating asymmetries for $B_q
\to \ks \ksb$ decays in the pQCD approach.

For the charged $B_u$ and $B_c$ decays, the direct CP-violating
asymmetry $\acp^{\rm dir}$ can be defined as:
 \beq
\acp^{\rm dir} =  \frac{|\overline{\cal M}|^2 - |{\cal M}|^2}{
 |\overline{\cal M}|^2+|{\cal M}|^2}\;,
\label{eq:acp1}
\eeq
where $M$ denotes the decay amplitude of charged $B_{u(c)} \to {\ks}^+ {\ksb}^0$ decays, while $\overline{M}$ stands for the charge conjugation one correspondingly.

Using Eq.~(\ref{eq:acp1}), it is easy to calculate the
direct CP-violating asymmetries as listed in Eq.~(\ref{eq:diru}) for the considered $B_u$
decay in S1 and S2,
\beq
{\cal A}_{CP}^{\rm dir}(B_u \to {\ks}^+ {\ksb}^0) &\approx& \left\{ \begin{array}{ll}
\hspace{0.35cm}33.7^{+0.2}_{-2.2}(\omega_{b})^{+0.6}_{-0.3}(\bar f_S)^{+3.5}_{-3.5}(B_{i}^{S})
^{+2.0}_{-1.6}({\rm CKM})   \% & \quad ({\rm S1}) \\
-24.4^{+1.3}_{-0.3}(\omega_{b})^{+0.3}_{-0.4}(\bar
f_S)^{+5.2}_{-6.9}(B_{i}^{S}) ^{+1.1}_{-1.6}({\rm CKM})   \% &
\quad ({\rm S2}) \\ \end{array} \right.  \label{eq:diru}\;,
\eeq
The large CP-violating asymmetries~(\ref{eq:diru}) plus large branching ratios~(\ref{eq:bru}) in
both scenarios are clearly measurable in the B factories and LHC
experiments. If these physical quantities could be tested at the
predicted level, it is doubtless that one can determine the better
scenario of $\ks$ meson and further understand the involved QCD
dynamics.

Because only the tree topology is involved,  there is no direct CP violation
in the considered $B_c$ decay mode, i.e.,
${\cal A}_{CP}^{\rm dir}(B_c \to {\ks}^+ {\ksb}^0) =0 $ in both scenarios.

As for the CP-violating asymmetries for the neutral decays
$B_{d(s)} \to \ks \ksb$, the effects of
$B_{d(s)}-\overline{B}_{d(s)}$ mixing should be considered. The
CP-violating asymmetries of $B_{d(s)}(\overline{B}_{d(s)}) \to
{\ks}^+ {\ks}^-, {\ks}^0 {\ksb}^0$ decays are time dependent and
can be defined as \beq \acp &\equiv& \frac{\Gamma\left
(\overline{B}_{d(s)}(\Delta t) \to f_{CP}\right) -
\Gamma\left(B_{d(s)}(\Delta t) \to f_{CP}\right )}{ \Gamma\left
(\overline{B}_{d(s)}(\Delta t) \to f_{CP}\right ) + \Gamma\left
(B_{d(s)}(\Delta t) \to f_{CP}\right ) }\non &=& \acp^{\rm dir}
\cos(\Delta m_{(s)}  \Delta t) + \acp^{\rm mix} \sin (\Delta m_{(s)} \Delta
t), \label{eq:acp-def} \eeq where $\Delta m_{(s)}$ is the mass
difference between the two $B_{d(s)}^0$ mass eigenstates, $\Delta
t =t_{CP}-t_{tag} $ is the time difference between the tagged
$B_{d(s)}^0$ ($\overline{B}_{d(s)}^0$) and the accompanying
$\overline{B}_{d(s)}^0$ ($B_{d(s)}^0$) with opposite $b$ flavor
decaying to the final CP-eigenstate $f_{CP}$ at the time $t_{CP}$.
The direct and mixing induced CP-violating asymmetries $\acp^{\rm
dir} ({\cal C}_f)$ (or ${\cal A}_f$ in term of Belle
Collaboration) and $\acp^{\rm mix} ({\cal S}_f)$ can be written as
\beq \acp^{\rm dir}= {\cal C}_f = \frac{ \left |
\lambda_{CP}\right |^2-1 } {1+|\lambda_{CP}|^2}, \qquad \acp^{\rm
mix}={\cal S}_f= \frac{ 2 {\rm Im}
(\lambda_{CP})}{1+|\lambda_{CP}|^2}, \label{eq:acp-csf} \eeq with
the CP-violating parameter $\lambda_{CP}$ \beq \lambda_{CP}
&\equiv& \eta_f \; \frac{V_{tb}^*V_{td(s)}}{V_{tb}V_{td(s)}^*}
\cdot \frac{ \langle f_{CP} |H_{eff}|\overline{B}_{d(s)}^0\rangle}
{\langle f_{CP} |H_{eff}|B_{d(s)}^0\rangle}. \label{eq:lambda2}
\eeq where $\eta_f$ is the CP-eigenvalue of the final states.
Moreover, for $B_s$ meson decays, a non-zero ratio $(\Delta
\Gamma/\Gamma)_{B_s}$ is expected in the
SM~\cite{Beneke99:Bsmixing,Fernandez06:Bsmixing}. For $B_s \to \ks
\ksb$ decays, the third term ${\cal A}_{\Delta \Gamma_s}$ related
to the presence of a non-negligible $\Delta \Gamma_s$
 to describe the CP violation can be defined as follows~\cite{Fernandez06:Bsmixing}:
\beq
{\cal A}_{\Delta \Gamma_s} &=& \frac{ 2 {\rm Re} ( \lambda_{CP})}{1+|\lambda_{CP}|^2},
\label{eq:acp-dgs}
\eeq
The three quantities describing the CP violation in $B_s$ meson decays shown in Eqs.~(\ref{eq:acp-csf}) and
(\ref{eq:acp-dgs}) satisfy the following relation,
\beq
 |\acp^{\rm dir}|^2+ |\acp^{\rm mix}|^2+ |{\cal A}_{\Delta \Gamma_s}|^2 &=&
 1 \;.\label{eq:summation-cp}
\eeq

For $B_{(s)}^0/\overline{B}_{(s)}^0 \to {\ks}^0 {\ksb}^0$ decays,
they do not exhibit CP violation in both scenarios,
since they involve only penguin
contributions at the leading order in the SM, as can be seen from the decay
amplitudes as given in Eqs.~(\ref{eq:tda-b2k0k0b}) and (\ref{eq:tda-bs2k0k0b}).
Then for $B_s^0/\overline{B}_s^0 \to
{\ks}^0 {\ksb}^0$ mode, according to Eq.~(\ref{eq:summation-cp}),
the third term of the CP violation ${\cal A}_{\Delta\Gamma_s} =  100\%$.
If the experimental data for the direct CP asymmetries
$\acp^{\rm dir}$ in $B_{d(s)} \to {\ks}^0 {\ksb}^0$ decays exhibit obviously nonzero,
which will indicate the existence of new physics beyond the SM and will provide a
very promising place to look for this exotic effect.

For $B_{(s)}^0/\overline{B}_{(s)}^0 \to {\ks}^+ {\ks}^-$ decays,
with the decay amplitudes shown in Eqs.~(\ref{eq:tda-b2kpkm}) and
(\ref{eq:tda-bs2kpkm}), we can find the numerical results in both
scenarios for the CP-violating asymmetries in the pQCD approach
are as follows,
 \beq
 {\cal A}_{CP}^{\rm dir}(B_d \to {\ks}^+
{\ks}^-) &\approx&  \left\{ \begin{array}{ll}
-64.9^{+0.0}_{-0.6}(\omega_{b})^{+0.8}_{-1.0}(\bar
f_S)^{+5.9}_{-3.7}(B_{i}^{S})
^{+4.3}_{-2.3}({\rm CKM})   \% & \quad ({\rm S1}) \\
\hspace{0.40cm}5.9^{+0.3}_{-0.2}(\omega_{b})^{+2.7}_{-2.3}(\bar
f_S)^{+11.2}_{-1.9}(B_{i}^{S}) ^{+0.2}_{-0.4}({\rm CKM})   \% &
\quad ({\rm S2}) \\ \end{array} \right.   \label{eq:dird2}  \;,
\eeq
 \beq
{\cal A}_{CP}^{\rm mix}(B_d \to {\ks}^+ {\ks}^-) &\approx&  \left\{ \begin{array}{ll}
-49.9^{+3.8}_{-6.9}(\omega_{b})^{+0.4}_{-0.5}(\bar f_S)^{+1.7}_{-4.8}(B_{i}^{S})
^{+9.0}_{-8.1}({\rm CKM})   \% & \quad ({\rm S1}) \\
-98.9^{+0.0}_{-0.2}(\omega_{b})^{+0.1}_{-0.1}(\bar f_S)^{+1.0}_{-0.9}(B_{i}^{S})
^{+1.1}_{-0.6}({\rm CKM})   \% & \quad ({\rm S2}) \\ \end{array} \right.   \label{eq:mixd2}  \;,
\eeq
 \beq
{\cal A}_{CP}^{\rm dir}(B_s \to {\ks}^+ {\ks}^-) &\approx&  \left\{ \begin{array}{ll}
\hspace{0.10cm}12.9^{+0.0}_{-0.1}(\omega_{bs})^{+0.1}_{-0.1}(\bar f_S)^{+1.6}_{-1.3}(B_{i}^{S})
^{+0.6}_{-0.8}({\rm CKM})   \% & \quad ({\rm S1}) \\
-3.2^{+0.4}_{-0.4}(\omega_{bs})^{+0.3}_{-0.2}(\bar f_S)^{+1.5}_{-1.6}(B_{i}^{S})
^{+0.2}_{-0.1}({\rm CKM})   \% & \quad ({\rm S2}) \\ \end{array} \right.   \label{eq:dirs2}  \;,
\eeq
 \beq
{\cal A}_{CP}^{\rm mix}(B_s \to {\ks}^+ {\ks}^-) &\approx&  \left\{ \begin{array}{ll}
3.0^{+1.6}_{-1.7}(\omega_{bs})^{+0.2}_{-0.1}(\bar f_S)^{+0.4}_{-0.1}(B_{i}^{S})
^{+0.1}_{-0.1}({\rm CKM})   \% & \quad ({\rm S1}) \\
1.8^{+0.2}_{-0.3}(\omega_{bs})^{+0.1}_{-0.1}(\bar f_S)^{+0.9}_{-1.9}(B_{i}^{S})
^{+0.1}_{-0.1}({\rm CKM})   \% & \quad ({\rm S2}) \\ \end{array} \right.   \label{eq:mixs2}  \;,
\eeq
 \beq
{\cal A}_{\Delta \Gamma_s}(B_s \to {\ks}^+ {\ks}^-) &\approx&  \left\{ \begin{array}{ll}
99.1^{+0.1}_{-0.0}(\omega_{bs})^{+0.0}_{-0.0}(\bar f_S)^{+0.2}_{-0.2}(B_{i}^{S})
^{+0.1}_{-0.1}({\rm CKM})   \% & \quad ({\rm S1}) \\
99.9^{+0.0}_{-0.0}(\omega_{bs})^{+0.0}_{-0.0}(\bar
f_S)^{+0.1}_{-0.0}(B_{i}^{S}) ^{+0.0}_{-0.0}({\rm CKM})   \% &
\quad ({\rm S2}) \\ \end{array} \right.   \label{eq:hf2}  \;,
\eeq
where the errors are induced by the uncertainties from shape
parameter $\omega_b$($\omega_{bs}$) for $B$($B_s$) meson, scalar
decay constant $\bar f_S$ and Gegenbauer moments $B_i^S$ in the
distribution amplitudes of scalar $\ks$, and CKM parameters ($\bar
\rho, \bar \eta$), respectively. For the direct CP asymmetries in
the $B_d \to {\ks}^+ {\ks}^-$ and $B_s \to {\ks}^+ {\ks}^-$ decays
for example, i.e., Eqs.~(\ref{eq:dird2}) and (\ref{eq:dirs2}), one
can find that their signs and magnitudes are rather different in
both scenarios within theoretical errors. In the former mode,
$\acp^{\rm dir}$ is ($-69.4 \sim -57.6$) in S1 and ($2.9 \sim
17.4$) in S2; while in the latter one, $\acp^{\rm dir}$ is ($11.4
\sim 14.6$) in S1 and ($-4.9 \sim -1.6$) in S2. But, it is clear to find that
the magnitudes of direct CP-violating asymmetries for these two decays
in S1 are much larger than those in S2 in the pQCD approach
correspondingly, which will be more easily tested by the ongoing LHC and
forthcoming SuperB experiments. Furthermore, once the predictions
on $\acp^{\rm dir}$ including both size and sign in S1 could be confirmed at the predicted
level by the stringent experimental measurements in the future,
which will be helpful to investigate the physical property of the
scalar $\ks$ and determine the better scenario describing its QCD
dynamics in turn, and vice versa. Meanwhile, the stringent test
of $\acp^{\rm dir} (B_d \to {\ks}^+ {\ks}^-)$ can also provide indirect evidences
for an important but
controversial issue on the evaluation of annihilation
contributions at leading power.

Finally, it is worthy of mentioning that we here just study the perturbative short-distance
contributions as the first step in present work. The large theoretical errors induced by the
large uncertainties of the inputs in the nonperturbative distribution amplitudes, such as
$\phi_{B_q}$, $\phi_{\ks}$, etc, should be constrained by the precision measurements, which
will be very helpful to explore the hadronic dynamics of $\ks$ and the QCD dynamics involved
in these considered decay channels. We do not consider the possible long-distance
contributions, such as the rescattering effects, although they should be present, and they
may be large and affect the theoretical predictions. It is beyond
the scope of this work and expected to be studied in the future.

\section{Summary}\label{sec:sum}

In this work, we studied the two-body charmless hadronic $B_q \to \ks \ksb (q= u, d, s, c)$
decays by employing the pQCD approach based on the
$k_T$ factorization theorem. Based on the assumption of
two-quark structure of the light scalar $\ks$, we made the theoretical predictions
and phenomenological analysis
on the physical observables: CP-averaged branching ratios and CP-violating asymmetries.

From our numerical evaluations and phenomenological analysis, we
found the following results:
\begin{itemize}
\item
In both scenarios of $\ks$,  $B_q \to \ks \ksb$ decays in the pQCD approach
exhibit the large CP-averaged branching ratios ($10^{-6} \sim 10^{-4}$),
which are clearly measurable in the present B factories and ongoing LHC experiments.

\item
In the considered six decay channels, only preliminary upper limit on
$Br(B_d^0 \to {\ks}^0 {\ksb}^0)$ mode has been reported.
The pQCD prediction basically agrees with this upper limit, and will be tested
when better experimental measurements become available.

\item
For $B_u \to {\ks}^+ {\ksb}^0$ decay, the large branching ratio plus large direct
CP-violating asymmetry will offer a great opportunity to test the hadronic dynamics
of $\ks$ and the QCD dynamics involved in the considered mode.

\item
The obvious nonzero $\acp^{\rm dir}$ for $B_{d(s)} \to {\ks}^0 {\ksb}^0$ decays will
provide a good platform to explore the exotic new physics effects beyond the SM.

\item
The pure annihilation process $B_c \to {\ks}^+ {\ksb}^0$ have a large branching ratio and
will be measured soon at the LHC experiments, which can help us to
understand the role of the annihilation contributions in B physics.

\item
The pQCD predictions still have large theoretical uncertainties induced by the
uncertainties of input parameters, such as the universal distribution amplitudes
$\phi_{B_q}$ and $\phi_{\ks}$, which should be well constrained by the precision data.

\end{itemize}

\begin{acknowledgments}

The authors would like to thank Cai-Dian L\"u and Hsiang-Nan~Li for helpful discussions.
X.~Liu thanks Y.M.~Wang for his comments.
This work is supported by the National Natural Science Foundation of China
under Grant No. 10975074, and No. 10735080, and by the Project on Graduate
Students' Education and Innovation of Jiangsu Province, under Grant No.
${\rm CX09B_{-}297Z}$, and by the Project on Excellent Ph.D Thesis of Nanjing Normal
University, under Grant No. 181200000251, and by the Project on Start-up Fund of Xuzhou Normal University.

\end{acknowledgments}



\end{document}